\documentclass[12pt]{article}
\usepackage{subeqn}
\usepackage{epsfig}
\newcommand{\be}{\begin{equation}}
\newcommand{\ee}{\end{equation}}

\newcommand{\no}{\noindent}
\newcommand{\ce}{\begin{center}}
\newcommand{\nc}{\end{center}}

\makeatletter
\@addtoreset{equation}{section}
\makeatother

\def\sqr#1#2{{\vcenter{\vbox{\hrule height.#2pt
\hbox{\vrule width.#2pt height#1pt \kern#1pt
\vrule width.#2pt} \hrule height.#2pt}}}}

\def\operp{\hbox{${\kern+.25em{\bigcirc}
\kern-.85em\bot\kern+.85em\kern-.25em}$}}

\def\lsim{\;\raise0.3ex\hbox{$<$\kern-0.75em\raise-1.1ex\hbox{$\sim$}}\;}
\def\gsim{\;\raise0.3ex\hbox{$>$\kern-0.75em\raise-1.1ex\hbox{$\sim$}}\;}

\def\no{\noindent}

\def\ce{\centerline}
\def\ve{\vfill\eject}
\def\rdots{\mathinner{\mkern1mu\raise1pt\vbox{\kern7pt\hbox{.}}\mkern2mu
\raise4pt\hbox{.}\mkern2mu\raise7pt\hbox{.}\mkern1mu}}

\def\Tr{{\rm Tr}}

\def\e e{$e^+ e^-$ }

\begin{document}

\ce{\bf A FIELD THEORY OF KNOTTED SOLITONS}

\vskip.3cm

\ce{\it Robert J. Finkelstein}
\vskip.3cm

\ce{Department of Physics and Astronomy}
\ce{University of California, Los Angeles, CA 90095-1547}

\vskip1.0cm

\no{\bf Abstract.} The conjecture that the elementary fermions are
knotted flux tubes permits the construction of a phenomenology that
is not accessible from the standard electroweak theory.  In order to
carry these ideas further we have attempted to formulate 
the elements of a field
theory in which local $SU(2)\times U(1)$, the symmetry group of
standard electroweak theory, is combined with global $SU_q(2)$, the
symmetry group of knotted solitons.

\vskip5.0cm

UCLA/07/TEP/2

\ve

\section{Introduction.}

We have previously discussed the phenomenology of 
$q$-solitons.$^{1,2,3}$  We should now like to incorporate the
$q$-soliton into a field theory that agrees closely with the
standard point particle electroweak theory and at the same time
replaces the point particles by knotted solitons.  To do this we
retain both local $SU(2)\times U(1)$, the symmetry group of
standard electroweak, and global $SU_q(2)$, the symmetry group of
the $q$-solitons.  This is done by retaining the formal structure
of standard electroweak but at the same time expanding the quantum
fields in a new class of normal modes determined by the
$SU_q(2)$ algebra.  Then all quantum vector fields will lie in
$SU(2)\times U(1)$ Lie algebra as well as in the disjoint $SU_q(2)$ algebra.

\vskip.5cm

\section{Lagrangians.}

The standard Lagrangian density is
\begin{eqnarray}
{\cal{L}} &=& -\frac{1}{4} (G^{\mu\nu}G_{\mu\nu} + H^{\mu\nu}H_{\mu\nu})
+ i[\bar L\not\nabla L + \bar R\not\nabla R] \nonumber \\
& &\mbox{}+ (\nabla\phi)^+\cdot\nabla\phi - V(\phi^+\phi) -
\frac{m}{\rho_0}(\bar L\phi R + \bar R\bar\phi^+L)
\end{eqnarray}
where
\begin{eqnarray*}
& &G^{\mu\nu} = \partial^\mu\vec W^\nu-\partial^\nu 
\vec W^\mu-g\vec W^\mu\times\vec W^\nu 
\hspace{.25in} R = \frac{1}{2}~(1+\gamma_5)\psi \\
& &H^{\mu\nu} = \partial^\mu W_o^\nu-\partial^\nu W_o^\mu
\hspace{.50in} L = \frac{1}{2}~(1-\gamma_5)\psi  \\
& &\nabla^\mu = \partial^\mu+i~g\vec W^\mu
\cdot \vec t + ig^\prime W_o^\mu t_o \\
\end{eqnarray*}
The changes that we make amount to primarily multiplying the usual normal
modes of $\vec W_\mu,\phi$, and $\psi$ by suitably chosen
$D^j_{m,m^\prime}(a,\bar a,b,\bar b)$,the elements of
the irreducible representations
of $SU_q(2)$. This step is analogous to appending the spin states to
$e^{i\vec p\vec x}$ and is determined by the symmetries of the
algebra of $(a,\bar a,b,\bar b)$.  We shall also introduce small
modifications of (2.1).

After the changes are made the new symmetry of the Lagrangian density
will be
\[
[\mbox{local}~SU(2)\times U(1)] \otimes [\mbox{global}~U(1)]
\]
and the new Lagrangian will be invariant under the new symmetry
group.

\vskip.5cm

\section{The Irreducible Representations of $SU_q(2)$.}

The $(2j+1)$-dimensional unitary irreducible representations of $SU_q(2)$ are
\be
D^j_{mm^\prime}(a,\bar a,b,\bar b) = \Delta^j_{mm^\prime} \sum_{s,t}
\left\langle\matrix{n_+ \cr s \cr}\right\rangle_1
\left\langle\matrix{n_- \cr t}\right\rangle_1
q_1^{t(n_++1-s)}(-1)^t\delta(s+t,n^\prime_+) 
a^sb^{n_+-s}\bar b^t\bar a^{n_--t} 
\ee
\no where
\be
\begin{array}{rcl}
n_\pm &=& j\pm m \\ n^\prime_\pm &=& j\pm m^\prime \\
\end{array} \quad
\left\langle\matrix{n \cr s \cr}\right\rangle_1 =
{\langle n\rangle_1!\over \langle s\rangle_1!\langle n-s\rangle_1!}
\quad \langle n\rangle_1 = {q_1^{2n}-1\over q_1^2-1} \nonumber
\ee
\no and  
\be
\Delta^j_{mm^\prime} = \left[{\langle n^\prime_+\rangle_1!~
\langle n^\prime_-\rangle_1!\over
\langle n_+\rangle_1!~\langle n_-\rangle_1!}\right]^{1/2} \qquad
q_1 = q^{-1} 
\ee
\no Here the arguments $(a,b,\bar a,\bar b)$ obey the following
algebra:
\begin{center}
\begin{tabular}{lllr}
$ab= qba$ & $a\bar a+b\bar b = 1$ & $b\bar b = \bar bb$ & \\
$a\bar b =q\bar ba$ & $\bar aa + q_1^2\bar bb = 1$ & & \hspace{1.25in}(A)
\end{tabular}
\end{center}
\no In the limit $q=1$, where $\langle n\rangle_1\to n$ and the
arguments commute, the $D^j_{mm^\prime}$ become the 
matrix elements of the irreducible
representations of $SU(2)$.  

\vskip.3cm

\no {\bf Remarks:}

Every term of (3.1) contains a product of 
non-commuting factors that may be reduced (after dropping numerical
factors) to the form
\be
a^{n_a}\bar a^{n_{\bar a}}b^{n_b}\bar b^{n_{\bar b}}
\ee
\no where $n_a = s$ and $n_{\bar a}=n_--t$.

But the factor $\delta(s+t,n^\prime_+)$ appearing in (3.1) implies
\be
n_+^\prime = n_a + (n_--n_{\bar a})
\ee
\no or
\be
n_a-n_{\bar a} = m+m^\prime
\ee
\no Note that (3.6) holds for \underline{all} terms of (3.1) 
and \underline{all} representations $(j)$.

By a similar argument one notes that
\be
n_b-n_{\bar b} = m-m^\prime
\ee
again holding for all terms of (3.1) and all representations.

\vskip.3cm

\no {\bf State Space:}

The $D^j_{mm^\prime}$ take on numerical values when allowed to
operate on a state space that may be defined as follows:

Since $[b,\bar b] = 0$, $b$ and $\bar b$ have common eigenstates.
Let $|0\rangle$ be the ground state and let
\begin{eqnarray}
b|0\rangle &=& \beta|0\rangle \\
\bar b|0\rangle &=& \beta^\star|0\rangle
\end{eqnarray}
From the algebra one finds
\be
\bar bb|n\rangle = E_n|n\rangle
\ee
where
\be
E_n = q^{2n}|\beta|^2
\ee
and
\be
|n\rangle \sim \bar a^n|0\rangle
\ee
Here $\bar bb$ is Hermitian with real eigenvalues and orthogonal 
eigenstates.

\vskip.5cm

\section{Invariance of the Algebra (A).}

The algebra (A) is invariant under the following transformations
\be
\begin{array}{rcl}
a^\prime &=& e^{i\varphi_a}a  \\
\bar a^\prime &=& e^{-i\varphi_a}\bar a
\end{array}
\ee
\be
\begin{array}{rcl}
b^\prime &=& e^{i\varphi_b}b  \\
\bar b^\prime &=& e^{-i\varphi_b}\bar b
\end{array}
\ee
The gauge transformations induced on the irreducible representations
are then
\begin{eqnarray}
U_aD^j_{mm^\prime} &=& e^{i\varphi_a(n_a-n_{\bar a})}D^j_{mm^\prime} \\
U_bD^j_{mm^\prime} &=& e^{i\varphi_b(n_b-n_{\bar b})}
D^j_{mm^\prime}
\end{eqnarray}
since (3.6) and (3.7) hold for all terms of $D^j_{mm^\prime}$.
Define with the aid of (3.6) and (3.7)
\begin{eqnarray}
Q_a &\equiv& k(n_a-n_{\bar a}) \equiv k(m+m^\prime) \\
Q_b &\equiv& k(n_b-n_{\bar b}) \equiv k(m-m^\prime)
\end{eqnarray}
Then (4.1) and (4.2) induce the following gauge transformations on
the irreducible representations:
\begin{eqnarray}
U_aD^j_{mm^\prime} &=& e^{iQ_a\varphi_a/k} D^j_{mm^\prime} \\
U_bD^j_{mm^\prime} &=& e^{iQ_b\varphi_b/k} D^j_{mm^\prime}
\end{eqnarray}
According to (4.5) and (4.6) both $Q_a$ and $Q_b$ reverse sign as
one goes from $D^j_{mm^\prime}$ to $\bar D^j_{mm^\prime}$.  Since we shall associate $D^j_{mm^\prime}$
and $\bar D^j_{mm^\prime}$ with particles and antiparticles,
$Q_a$ and $Q_b$ reverse sign as one passes to the antiparticle.

\vskip.5cm

\section{The Knot Representation.}

Since $SU_q(2)$ underlies the description of knots, we shall make this
connection explicit by labelling a knot with $D^{N/2}_{\frac{w}{2}
\frac{r+1}{2}}$ where $(N,w,r)$ are the number of crossings, the 
writhe, and the rotation of the knot.  The minimum value of $N$ is
3, and the corresponding knot is a trefoil.  There are four possible
trefoils and they are characterized by $(w,r) = (-3,2), (3,2),
(3,-2), (-3,-2)$.

We base our work on the physical conjecture that the elementary
fermions are knotted flux tubes.  More precisely we associate each 
of the four families of elementary fermions (leptons, neutrinos,
up quarks, down quarks) with one of the four trefoils and each of the
three members of a family with the three lowest states of excitation
of its trefoil.  Thus the electron, muon, and tau are the three
lowest states of the leptonic soliton.  Then the fermionic families
are also labelled by $D^{3/2}_{mm^\prime} = D^{3/2}_{\frac{w}{2}
\frac{r+1}{2}}$ and we shall relate $Q_a$ and $Q_b$ as defined in
(4.5) and (4.6), to the electric charge and hypercharge of the
corresponding family.  In fact we may match the 4 trefoils with the
4 families in Table 1 if and only if $k = -1/3$ in (4.5).

\vskip.3cm

\begin{center}
{\bf Table 1.}
\end{center}
\[
\begin{array}{ccrc}
\underline{(w,r)} \quad & \underline{D^{3/2}_{\frac{w}{2}\frac{r+1}{2}}} \quad & \underline{Q_a} \quad & \underline{\rm Family} \\
(-3,2) \quad & D^{3/2}_{-\frac{3}{2}\frac{3}{2}} \quad & 0
\quad & (\nu_e,\nu_\mu,\nu_\tau) \\
(3,2) \quad & D^{3/2}_{\frac{3}{2}\frac{3}{2}} \quad & -1 \quad &
(e,\mu,\tau) \\
(3,-2) \quad & D^{3/2}_{\frac{3}{2}-\frac{1}{2}} \quad & -\frac{1}{3}
\quad & (dsb) \\
(-3,-2) \quad & D^{3/2}_{-\frac{3}{2}-\frac{1}{2}} \quad & \frac{2}{3}
\quad & (uct) \\
\end{array}
\]

\vskip.3cm

\no Here
\be
Q_a = -\frac{1}{3} (m+m^\prime)
\ee
Note that $m$ distinguishes between the two members of a doublet
while $m^\prime$ labels either the light $(e,\nu)$ doublets or the
heavy $(u,d)$ doublet.
To complete the correspondence we write $D^j_{mm^\prime}$ in terms
of the conventional quantum numbers that label the point particles of
standard theory.  Since all fermions have $t = 1/2$ and $t_3 = \pm 1/2$
while the trefoils have $N=3$ and $w=\pm 3$ we may set
\begin{eqnarray}
t &=& \frac{N}{6} \\
t_3 &=& -\frac{w}{6}
\end{eqnarray}
or
\begin{subequations}
\begin{eqnarray}
D^j_{mm^\prime} &=& D^{N/2}_{\frac{w}{2}\frac{r+1}{2}} \\
&=& D^{3t}_{-3t_3~m^\prime} 
\end{eqnarray}
\end{subequations}
But
\be
\begin{array}{rcl}
Q_a &=& -\frac{1}{3} (m+m^\prime) \\
&=& t_3 - \frac{m^\prime}{3}
\end{array}
\ee
Since $Q_a$ is the electric charge, we also have according to the
standard theory:
\be
Q_a = t_3 + t_0
\ee
\no so that by (5.5)
\be
t_0 = -\frac{m^\prime}{3}
\ee
and
\be
Q_b = t_3-t_0
\ee
Hence the two charges $Q_a$ and $Q_b$ stemming from the gauge
invariance of the internal algebra (A) are simply related to the
electric charge and hypercharge appearing in standard theory.
We also note that $Q_a=0$ implies $n_a=n_{\bar a}$.  Then $a$ and
$\bar a$ may be eliminated from (3.4) since
\[
a^n\bar a^n = \prod^{n-1}_{s=0} (1-q^{2s}b\bar b)
\]
Therefore neutral states (neutrinos and neutral bosons) lie entirely
in the $(b,\bar b)$ subalgebra.

Since these transformations on the internal algebra are assumed to be
global, they do not give rise to new fields.  Rather they simply
define the charge and hypercharge of the fermionic sources of the
vector fields.

Having found the explicit representations $D^{3/2}_{mm^\prime}$
of the fermions we now turn to the representations of the vector
bosons.  Since the vector bosons are responsible for pair production,
we shall represent them by ditrefoils with $N=6$.  Then by (5.2) and 
(5.4a)
\begin{eqnarray}
t &=& \frac{N}{6} = 1 \\
j &=& \frac{N}{2} = 3
\end{eqnarray}
We now find Table 2
\begin{center}
{\bf Table 2.}
\end{center}
\[
\begin{array}{lcrcll}
& ~~~ \underline{t} ~~& \underline{t_3} ~~& \underline{t_0} &
\underline{D^3_{mm^\prime}}  \\
W^+ & ~~~1 ~~& 1 ~~& 0 ~~~& D^3_{-30}  \\
W^- & ~~~1 ~~& -1 ~~& 0 ~~& D^3_{30}  \\
W^3 & ~~~1 ~~& 0 ~~& 0 ~~& D^3_{00}  \\
\end{array}
\]
where $t_3$ and $t_0$ are taken from standard electroweak theory.
Here we have used
\begin{eqnarray}
m &=& -3t_3 \\
m^\prime &=& -3t_0
\end{eqnarray}
\no that follow from (5.4) and (5.7).

Since $W^0$ in the standard theory arises from coupling to $U(1)$ it
does not belong to the isotopic spin multiplet, $t=1$, and
therefore its representation cannot be fixed by $t_3$ and $t_0$.  Since
$Q = -\frac{1}{3}(m+m^\prime)$ represents the electric charge, all
that can be said about $W^0$ at this point is that $m+m^\prime=0$.
$W^0$ can then be represented by any $D^j_{mm^\prime}$, where
$m+m^\prime=0$, except $D^3_{00}$, which is excluded unless $W^0$ is
otherwise distinguished from $W^3$.

Dropping numerical factors, the explicit monomials that label the
four fermionic solitons are as follows:
\be
D_\nu^{3/2}=\bar b^3, \quad D_\ell^{3/2}= a^3, \quad
D_u^{3/2}=\bar a^2\bar b, \quad D_d^{3/2}= ab^2
\ee
\no The corresponding forms labelling the charged vectors are
\begin{subequations}
\be
{\cal{D}}_-^3= a^3b^3,\quad {\cal{D}}_+^3=\bar b^3\bar a^3
\ee
\no while ${\cal{D}}_3^3$ and ${\cal{D}}_o^3$, labelling the neutral vectors are
polynomials lying in the $(b,\bar b)$ subalgebra.  One possibility
is
\be
{\cal{D}}_3^3= D_{oo}^3 \qquad {\cal{D}}^3_0= D_{-11}^3
\ee
\end{subequations}

\vskip.5cm

\section{The Normal Modes of the Quantum Fields.}

In the proposed formalism the normal modes of the standard electroweak
vector fields are to be modified by multiplying the generators
$(\vec t,t_o)$ of the Lie algebra of $SU(2)\times U(1)$ by the
${\cal{D}}_\alpha^3(a,\bar a,b,\bar b)$,~$\alpha = (+,-,3,0)$ as follows:
\begin{subequations} 
\begin{eqnarray}
\vec t &\to& \vec \tau = (c_+t_+{\cal{D}}_+^3,c_-t_-{\cal{D}}_-^3,
c_3t_3{\cal{D}}_3^3) \\
t_0 &\to& \tau_0 = t_0{\cal{D}}_0^3
\end{eqnarray}
\end{subequations}
In the fundamental representation
\be
(\vec\tau,\tau_0) = \left(
\begin{array}{cc}
0 & c_+{\cal{D}}_+^3 \\ 0 & 0 
\end{array} \right)~, ~ \left(
\begin{array}{cc}
0 & 0 \\ c_-{\cal{D}}_-^3 & 0 
\end{array} \right)~, ~ \left(
\begin{array}{cc}
c_3{\cal{D}}_3^3 & 0 \\ 0 & -c_3{\cal{D}}_3^3
\end{array} \right)~, ~ \left(
\begin{array}{cc}
c_0{\cal{D}}_0^3 & 0 \\ 0 & c_0{\cal{D}}_0^3
\end{array} \right)
\ee
where the ${\cal{D}}_\alpha^3~(\alpha = +,-,3,0)$ are given by (5.14).  The
numerically valued matrices $(\vec t,t_0)$ are in this way replaced
by the operator valued $(\vec\tau,\tau_0)$.

For the fermions and Higgs-like fields one replaces the numerically
valued 2-rowed spinors of isotopic spin $SU(2)$ by the following
operator valued spinors
\begin{eqnarray}
& &\left( 
\begin{array}{c}
D_\nu^{3/2} \\ 0 
\end{array} \right),~ \left(
\begin{array}{c}
0 \\ D_\ell^{3/2}
\end{array} \right),~\left(
\begin{array}{c}
D_u^{3/2} \\ 0
\end{array} \right),~\left(
\begin{array}{c}
0 \\ D_d^{3/2}
\end{array} \right) \nonumber \\
& &\mbox{}= \left(
\begin{array}{c}
1 \\ 0
\end{array} \right)~D_\nu^{3/2},~ \left(
\begin{array}{c}
0 \\ 1 
\end{array} \right)~D_\ell^{3/2},~ \left(
\begin{array}{c}
1 \\ 0
\end{array} \right)~D_u^{3/2},~ \left(
\begin{array}{c}
0 \\ 1
\end{array} \right)~D_d^{3/2}
\end{eqnarray}
where $D^{3/2}_r$ is an abbreviation for the irreducible representation
associated with the $r^{\rm th}$ soliton and where 
$r = (\nu,\ell,u,d)$.
We also introduce $\psi_{Ari}$
\begin{eqnarray}
\psi_{1ri} &=& \psi_{1r}~D_r(a\bar a b\bar b)|i\rangle \quad
A=1 \quad r=\nu,\ell \quad i=1,2,3 \nonumber \\
\psi_{2ri} &=& \psi_{2r}~D_r(a\bar a b\bar b)|i\rangle \quad
A=2 \quad r=u,d \quad i=1,2,3
\end{eqnarray}
where the lepton and neutrino solitons are combined into one isotopic
spinor, $\psi_{1ri}$, and the up and down quarks into a second isotopic
spinor, $\psi_{2ri}$ and where $i$ runs over the three states of the
soliton.  Then
\be
\psi_1 = \left(
\begin{array}{l}
\psi_\nu ~D^{3/2}_\nu|i\rangle \\
\psi_\ell ~D^{3/2}_\ell|i\rangle
\end{array} \right) \quad \mbox{and} \quad
\psi_2 = \left(
\begin{array}{l}
\psi_u ~D^{3/2}_u|i\rangle \\
\psi_d ~D^{3/2}_d|i\rangle
\end{array} \right)
\ee
We introduce the 4 bosons by defining
\be
\not{\cal{W}}_{rs} = ig\not\vec W\vec\tau_{rs} + ig^\prime \not W^0(\tau_0)_{rs}
\ee
where
\begin{eqnarray}
(\tau_\pm)_{rs} &=& c_\pm(t_\pm)_{rs} {\cal{D}}_\pm^3 \nonumber \\
(\tau_3)_{rs} &=& c_3(t_3)_{rs} {\cal{D}}^3_3 \nonumber \\
(\tau_0)_{rs} &=& c_0(t_0)_{rs} {\cal{D}}^3_0
\end{eqnarray}
and $(\vec W,W^0)$ are numerically valued and replace the components 
of the standard boson field and
${\cal{W}}$ lies in the internal algebra.  The numerical coefficients $(c_\pm,c_3,c_0)$ in (6.7) are to be considered
functions of $q$ and $\beta$, the parameters of the model. They are
fixed by the relative masses of the vectors and will be so
determined in Section 10. 

Here $({\cal{D}}^3_\pm,
{\cal{D}}^3_3,
{\cal{D}}^3_0)$ are given by (3.1) and (5.14).  The covariant derivative is now
\be
\not\nabla_{rs} = \delta_{rs}\not\partial + \not{\cal{W}}_{rs}
\ee
The corresponding field strengths are
\be
{\cal{W}}_{\mu\lambda} = [\nabla_\mu,\nabla_\lambda]
\ee
We shall introduce the direct boson-fermion interactions as follows:
\be
(\bar\psi_A)_{ri}\not\nabla_{rs}U_A(\psi_A)_{si^\prime} \quad
A = 1,2 \quad (r,s) = (\nu,\ell)~\mbox{or}~
(u,d)~\mbox{and}~~ i,i^\prime = 1,2,3
\ee
where $A=1$ labels the $(\nu,\ell)$ doublet and $A=2$ labels the quark
doublet $(u,d)$ according to (6.4).

The form of $U_1$ is restricted by the universal Fermi interaction,
while $U_2$ replaces the Kobayashi-Maskawa matrix.  Here $U_2$
``rotates" the initial state $(\psi_A)_{si^\prime}$.

\vskip.5cm

\section{The $\tau$-Commutators.}

The total field strength is by (6.9) and (6.6)
\begin{eqnarray}
{\cal{W}}_{\mu\lambda} &=& [\nabla_\mu,\nabla_\lambda]\\ \nonumber
&=& ig(\partial_\mu\vec W_\lambda-\partial_\lambda\vec W_\mu)\vec\tau
+ ig^\prime(\partial_\mu W_\lambda^0-\partial_\lambda W_\mu^0)
\tau_0-g^2W_\mu^m W_\lambda^k[\tau_m,\tau_k]
\end{eqnarray}

In extracting the $\tau_k$-independent fields from the Tr 
$\bar{\cal{W}}^{\mu\lambda}{\cal{W}}_{\mu\lambda}$ invariant one needs
to compute commutators of the $\tau_k$.  By (6.7)
\be
[\tau_k,\tau_\ell] = c_kc_\ell[t_k{\cal{D}}_k,t_\ell{\cal{D}}_\ell]
\ee
where
\[
t_k = (t_+,t_-,t_3)
\]
or
\be
t_k = \left(
\begin{array}{cc}
0 & 1 \\ 0 & 0
\end{array} \right)~, \left(
\begin{array}{cc}
0 & 0 \\ 1 & 0
\end{array} \right)~, \left(
\begin{array}{cr}
1 & 0 \\ 0 & -1
\end{array} \right)
\ee
and
\be
{\cal{D}}_k = ({\cal{D}}_+,{\cal{D}}_-.{\cal{D}}_3)
\ee
where by (5.14)
\begin{eqnarray}
{\cal{D}}_+ &=& D^3_{-30}/N_+ = \bar b^3\bar a^3 \\
{\cal{D}}_- &=& D^3_{30}/N_- = a^3b^3 \\
{\cal{D}}_3 &=& D^3_{00} = f_3(b\bar b)
\end{eqnarray}
where the $N_\pm$ are $c$-number normalizing factors
and $f_3$ is a function of $b\bar b$ as determined by (3.1).
Note
\be
\bar{\cal{D}}_+ = {\cal{D}}_- \qquad \mbox{and} \qquad
\bar{\cal{D}}_3 = {\cal{D}}_3
\ee
One has the familiar
\be
\begin{array}{rcl}
\left[t_+,t_3\right] &=& -2t_+ \\
\left[t_-,t_3\right] &=& 2t_- \\
\left[t_+,t_-\right] &=& t_3
\end{array}
\ee
and
\be
[t_i,t_j] = c_{ij}^kt_k \qquad (i,j,k) = (+,-,3)
\ee
Similarly one has
\begin{eqnarray}
\left[{\cal{D}}_+,{\cal{D}}_3\right] &=&
\hat c_{+3}^+{\cal{D}}_+ \\
\left[{\cal{D}}_-,{\cal{D}}_3\right] &=&
\hat c_{-3}^-{\cal{D}}_- \\
\left[{\cal{D}}_+,{\cal{D}}_-\right] &=&
\hat c_{+-}^3{\cal{D}}_3
\end{eqnarray}
where
\be
\hat c_{+3}^+ = f_3(q_1^6\bar bb)-f_3(\bar bb)
\ee
By (7.8) and (7.11)
\be
\hat c_{-3}^- = -\hat c_{+3}^+ 
\ee
and by (7.5)-(7.7)
\be
\hat c_{+-}^3 = (b\bar b)^3[\bar a^3a^3-q^{18}a^3\bar a^3]/f_3(\bar bb)
\ee
Here
\begin{eqnarray}
\bar a^3a^3 &=& \prod_{t=1}^3 (1-q_1^{2t}b\bar b) \\
a^3\bar a^3 &=&\prod_{t=0}^2 (1-q^{2t}b\bar b)
\end{eqnarray}
It follows from these equations and from (7.14)-(7.16) that all
$\hat c_{ij}^k$ are functions of $b\bar b$ and mutually commute.

We now return to (7.2).  Since $t_k$ and ${\cal{D}}_\ell$ commute,
we have
\be
[\tau_k,\tau_\ell] = c_kc_\ell([t_k,t_\ell]{\cal{D}}_k{\cal{D}}_\ell
+t_\ell t_k[{\cal{D}}_k,{\cal{D}}_\ell])
\ee
By (7.3)
\begin{eqnarray}
& &t_+t_3 = -t_+  \qquad\qquad t_3t_+ = t_+ \\
& &t_-t_3 = t_- \qquad\qquad ~~t_3t_- = -t_- \\
& &t_+t_- = \frac{1}{2} t_3 + \frac{1}{2}  \\
& &t_-t_+ = -\frac{1}{2} t_3 + \frac{1}{2}  
\end{eqnarray}
By (7.5)-(7.7)
\begin{eqnarray}
{\cal{D}}_+{\cal{D}}_3 &=& \bar b^3\bar a^3f_3(b\bar b) = f_3
(q_1^6b\bar b){\cal{D}}_+ \\
{\cal{D}}_-{\cal{D}}_3 &=& a^3b^3f_3(b\bar b) = f_3(q^6b\bar b)
{\cal{D}}_- \\
{\cal{D}}_+{\cal{D}}_- &=& (\bar b^3\bar a^3)(a^3b^3) = (\bar bb)^3
(\bar a^3a^3) = f_{+-}(\bar bb){\cal{D}}_3 \\
{\cal{D}}_-{\cal{D}}_+ &=& (a^3b^3)(\bar b^3\bar a^3) =
(\bar bb)^3 q^{18}(a^3\bar a^3) = f_{-+}(\bar bb){\cal{D}}_3
\end{eqnarray}

Here $f_{\pm\mp}$ and $f_3$ are all functions of $\bar bb$ and
therefore commute since $\bar a^3a^3,a^3\bar a^3$, and
${\cal{D}}_3$ are all functions of $\bar bb$.

The parts of (7.19) are then by (7.9), and by (7.20)-(7.27),
\begin{eqnarray}
& &[t_k,t_\ell] = c^s_{k\ell}t_s \hspace{1.5cm}
t_kt_\ell = \gamma^s_{k\ell}t_s + \gamma_{k\ell} \\
& &[{\cal{D}}_k,{\cal{D}}_\ell] = \hat c^s_{k\ell}{\cal{D}}_s \hspace{1.0cm}
{\cal{D}}_k{\cal{D}}_\ell = \hat\gamma^s_{k\ell}{\cal{D}}_s
\end{eqnarray}
where
\be
\gamma_{k\ell} = \frac{1}{2} \delta(k,\pm) \delta(\ell,\mp)
\ee
The coefficients $c^s_{k\ell}$ and $\gamma^s_{k\ell}$ are
numerically valued while $\hat c^s_{k\ell}$ 
and $\hat\gamma^s_{k\ell}$ are functions of $\bar bb$.

By (7.19) and (7.28)-(7.30), one now has
\be
[\tau_k,\tau_\ell] = c_kc_\ell[(c^s_{k\ell}t_s)
(\hat\gamma^s_{k\ell}{\cal{D}}_s) + (\gamma^s_{\ell k}t_s +
\gamma_{\ell k})(\hat c^s_{k\ell}{\cal{D}}_s)]
\ee
or
\be
[\tau_k,\tau_\ell] = \frac{c_kc_\ell}{c_s}
C^s_{k\ell}\tau_s + c_kc_\ell\gamma_{\ell k}\hat c^s_{k\ell}{\cal{D}}_s
\ee
where
\be
C^s_{k\ell} = c^s_{k\ell}\hat\gamma^s_{k\ell} +
\gamma_{\ell k}\hat c^s_{k\ell}
\ee
The structure coefficients of the algebra (7.32) and (7.33) depend
on the numerically valued $c^s_{k\ell}$ and $\gamma^s_{\ell k}$
but also on the $\hat c^s_{k\ell}$ and $\hat\gamma^s_{k\ell}$
that are functions of $\bar bb$ and are therefore also numerically
valued when allowed to operate on the states $|n\rangle$ of the
$q$-oscillator.

\vskip.5cm

\section{The Field Invariant $\langle 0|{\rm Tr}~{\cal{W}}_{\mu\lambda}{\cal{W}}^{\mu\lambda}|0\rangle$.}

Here the trace is taken on only the matrix part that is dependent on the $t$.
By (7.1) the non-Abelian contribution to the field strength is
\begin{eqnarray}
{\cal{W}}_{\mu\lambda} &=& ig(\partial_\mu W^s_\lambda-
\partial_\lambda W^s_\mu)\tau_s-g^2W_\mu^mW_\lambda^\ell
[\tau_m,\tau_\ell] \\
&=& ig(\partial_\mu W^s_\lambda-\partial_\lambda W^s_\mu)
\tau_s-g^2[c_mc_\ell c^{-1}_sC^s_{m\ell}\tau_s+c_mc_\ell\gamma_{\ell m}\hat c^s_{m\ell}{\cal{D}}_s]W_\mu^mW_\lambda^\ell
\end{eqnarray}
By (7.32), or
\be
{\cal{W}}_{\mu\lambda} = W^s_{\mu\lambda}\tau_s + \hat W^s_{\mu\lambda} {\cal{D}}_s
\ee
where
\begin{eqnarray}
W^s_{\mu\lambda} &=& ig(\partial_\mu W^s_\lambda-\partial_\lambda
W^s_\mu)-g^2c_mc_\ell c_s^{-1}C^s_{m\ell}W^m_\mu W^\ell_\lambda \\
\hat W^s_{\mu\lambda} &=& -\frac{1}{2}g^2c_mc_\ell\delta(\ell,\pm)
\delta(m,\mp)\hat c^s_{m\ell}W^m_\mu W^\ell_\lambda
\end{eqnarray}
by (7.30).  Then
\begin{eqnarray}
{\cal{W}}_{\mu\lambda}{\cal{W}}^{\mu\lambda} &=&
W^s_{\mu\lambda}W^{r\mu\lambda}\tau_s\tau_r+\hat W^s_{\mu\lambda}
\hat W^{\mu\lambda r}{\cal{D}}_s{\cal{D}}_r \nonumber \\
& & \mbox{} +W^s_{\mu\lambda}\tau_s\cdot\hat W^{r\mu\lambda}
D_r + \hat W^s_{\mu\lambda}{\cal{D}}_s\cdot W^{r\mu\lambda}\tau_r
\end{eqnarray}

In order to reduce the invariant ${\rm Tr}\langle 
0|{\cal{W}}_{\mu\lambda}{\cal{W}}^{\mu\lambda}|0\rangle$ we next
consider the expectation value of the expression (8.6) on the 
state $|0\rangle$
\begin{eqnarray}
\langle 0|{\cal{W}}_{\mu\lambda}{\cal{W}}^{\mu\lambda}|0\rangle &=&
\langle 0|W^s_{\mu\lambda}W^{r\mu\lambda}\tau_s\tau_r +
\hat W^s_{\mu\lambda}\hat W^{\mu\lambda r}{\cal{D}}_s{\cal{D}}_r 
\nonumber \\
& &\mbox{} +(W^s_{\mu\lambda}\tau_s)\cdot(\hat W^{\mu\lambda r}
{\cal{D}}_r) + (\hat W^s_{\mu\lambda}{\cal{D}}_s)\cdot
(W^{\mu\lambda r}\tau_r)|0\rangle
\end{eqnarray}
where ${\cal{W}}_{\mu\lambda}$ is given by (8.3)
\begin{eqnarray}
&=& \sum_n\left[\langle 0|W^s_{\mu\lambda}W^{r\mu\lambda}|n\rangle
\langle n|\tau_s\tau_r|0\rangle + \langle 0|\hat W^s_{\mu\lambda}
\hat W^{\mu\lambda r}|n\rangle\langle n|{\cal{D}}_s
{\cal{D}}_r|0\rangle\right] \nonumber \\
& &\mbox{}+\sum_n\left[\langle 0|W^s_{\mu\lambda}\hat W^{\mu\lambda r}|n\rangle\langle n|\tau_s{\cal{D}}_r|0\rangle +
\langle 0|\hat W^r_{\mu\lambda}W^{\mu\lambda s}|n\rangle
\langle n|{\cal{D}}_r\tau_s|0\rangle\right]
\end{eqnarray}
where $|n\rangle$ are the states of the $q$-oscillator.

Since $W^s_{\mu\lambda}$ and $\hat W^s_{\mu\lambda}$ depend on the
algebra only through $\bar bb$, they have no off-diagonal elements in
$n$.  Then
\begin{eqnarray}
\langle 0|{\cal{W}}_{\mu\lambda}{\cal{W}}^{\mu\lambda}|0\rangle &=&
\langle 0|W^s_{\mu\lambda}W^{r\mu\lambda}|0\rangle
\langle 0|\tau_s\tau_r|0\rangle + \langle 0|\hat W^s_{\mu\lambda}
\hat W^{\mu\lambda r}|0\rangle\langle 0|{\cal{D}}_s
{\cal{D}}_r|0\rangle \nonumber \\
& &\mbox{} +\langle 0|W^s_{\mu\lambda}\hat W^{r\mu\lambda}|0\rangle
\langle 0|\tau_s{\cal{D}}_r|0\rangle + \langle 0|\hat W^r_{\mu\lambda} W^{\mu\lambda s}|0\rangle
\langle 0|{\cal{D}}_r\tau_s|0\rangle
\end{eqnarray}

To continue the reduction of the field invariant
$I = \langle 0|{\rm Tr}~
{\cal{W}}_{\mu\lambda}{\cal{W}}^{\mu\lambda}|0\rangle$ we next
compute
\begin{eqnarray}
\langle 0|{\rm Tr}~\tau_s\tau_r|0\rangle &=&
c_sc_r\langle 0|({\rm Tr}~t_st_r){\cal{D}}_s{\cal{D}}_r|0\rangle
\nonumber \\
&=& c_sc_r({\rm Tr}~t_st_r)\langle 0|{\cal{D}}_s{\cal{D}}_r|0\rangle
\end{eqnarray}
since the trace is taken only on matrices dependent on the $t$. Then by (7.3)
\be
{\rm Tr}~t_st_r = \delta(s,\pm)\delta(r,\mp)+2\delta(s,3)
\delta(r,3)
\ee
One also finds by (7.5)-(7.7)
\be
\langle 0|{\cal{D}}_s{\cal{D}}_r|0\rangle =
[\delta(s,\pm)\delta(r,\mp)
+\delta(s,3)\delta(r,3)]\langle 0|
\bar{\cal{D}}_s{\cal{D}}_s|0\rangle
\ee
while
\be
\langle 0|{\rm Tr}~\tau_s{\cal{D}}_r|0\rangle =
\langle 0|{\rm Tr}~{\cal{D}}_r\tau_s|0\rangle = 0
\ee
Then the field invariant reduces to the following expression
\be
I = \sum_{s,r=(+-)}
\langle 0|A_{sr}W_{\mu\lambda}^{~~s}W^{\mu\lambda r} +
2\hat W_{\mu\lambda}^{~~s}\hat W^{\mu\lambda r}|0\rangle
\langle 0|{\cal{D}}_s{\cal{D}}_r|0\rangle
\ee
where
\be
A_{sr} = c_sc_r~{Tr}~t_st_r
\ee
Here $\langle 0|{\cal{D}}_s{\cal{D}}_r|0\rangle = 0$ unless
${\cal{D}}_s=\bar{\cal{D}}_r$ and
\begin{eqnarray}
& &\langle 0|\bar{\cal{D}}_+{\cal{D}}_+|0\rangle =
\langle 0|a^3b^3\bar b^3\bar a^3|0\rangle = q^{18}
\langle 0|(\bar bb)^3a^3\bar a^3|0\rangle \\
& &\langle 0|\bar{\cal{D}}_-{\cal{D}}_-|0\rangle =
\langle 0|\bar b^3\bar a^3a^3b^3|0\rangle =
\langle 0|\bar bb)^3\bar a^3a^3|0\rangle \\
& &\langle 0|\bar{\cal{D}}_3{\cal{D}}_3|0\rangle =
|f(\bar bb)|^2
\end{eqnarray}
These matrix elements are all functions of $\bar bb$, while
$W^s_{\mu\lambda}$ and $\hat W^s_{\mu\lambda}$ 
in (8.14) are given by (8.4)
and (8.5) respectively. 

The expression $W^s_{\mu\lambda}$ is of the same form as in the
standard theory but the structure coefficients differ from those
of the SU(2) algebra because they depend on $\bar bb$.
Since (8.14) is evaluated on the state $|0\rangle$ all expressions
of the form $F(\bar bb)$ become $F(|\beta|^2)$.  Therefore
the structure constants $C^s_{m\ell}(\bar bb)$ buried in 
$W^s_{\mu\lambda}$ and in turn appearing in (8.14) become
$C^s_{m\ell}(|\beta|^2)$.  Then the final reduced form of
$\langle 0|{\rm Tr}~{\cal{W}}_{\mu\lambda}{\cal{W}}^{\mu\lambda}
|0\rangle$ will have one part $(W^s_{\mu\lambda}W_s^{\mu\lambda})$
essentially the same as the standard theory, but with structure
constants $C^s_{m\ell}$ depending on $|\beta|^2$.  There is also a
second part $(\hat W^s_{\mu\lambda}\hat W_s^{\mu\lambda})$
given in (8.5) depending on $\hat c^s_{m\ell}$ which is (by
(7.14)-(7.16)) dependent on $q$ and $\beta$.  The sum of these two
parts is multiplied by $\langle 0|\bar{\cal{D}}_s{\cal{D}}_s|0
\rangle$, again a function of $q$ and $\beta$ as shown in (8.16)-
(8.18) and (7.17) and (7.18).  The functions $c_s(q,\beta)$ will be
given in Section 10.

\vskip.5cm

\section{Gauge Invariance.}

The new gauge group is generated by the following 
unitary transformations:
\be
{\cal{S}} = S\otimes s
\ee
where
\be
S \in \mbox{local}[SU(2)\otimes U(1)]
\ee
and
\be
s \in \mbox{global}~ U_a(1)\otimes U_b(1)
\ee
or
\be
S = e^{i\vec t\vec\theta(x)}e^{it_0\theta_0(x)}
\ee
and
\be
s = e^{iQ_a\theta_a}e^{iQ_b\theta_b}
\ee
where $\theta_a$ and $\theta_b$ are independent of $x$.
Then
\begin{eqnarray}
{\cal{S}}\psi_1 &=& {\cal{S}}\left(
\begin{array}{c}
D_\nu \\ D_\ell 
\end{array} \right) = S \otimes s\left(
\begin{array}{c}
D_\nu \\ D_\ell
\end{array} \right) \\
&=& S\left(
\begin{array}{c}
D_\nu^\prime \\ D_\ell^\prime
\end{array} \right) \\
&=& e^{i\vec t\vec\theta(x)}e^{it_0\theta_0(x)}\left(
\begin{array}{c}
D_\nu^\prime \\ D_\ell^\prime
\end{array} \right)
\end{eqnarray}
where
\be
D_k^\prime = e^{iQ_a(k)\theta_a}e^{iQ_b(k)\theta_b}D_k \qquad
k = (\nu,\ell)
\ee
with similar equations for the quark solitons.
Eq. (9.5) describes the action of $s$ on irreducible representations
only.  In general however $s$ denotes the action of the gauge 
transformations (4.1) and (4.2) on the entire Lagrangian including
the interaction (6.10).

In (6.10) therefore
\begin{eqnarray*}
s~U_2\psi_2 &=& U_2^\prime\psi_2^\prime \\
&=& U_2^\prime \left(
\begin{array}{c}
D_u^\prime \\ D_d^\prime
\end{array} \right)
\end{eqnarray*}
The interaction terms will transform as
\be
(\bar\psi_A)^\prime\not\nabla^\prime (U_A\psi_A)^\prime =
\bar\psi_A\bar{\cal{S}}\not\nabla^\prime {\cal{S}} (U_A\psi_A) 
\ee
Since ${\cal{S}}$ is unitary 
\be
{} = \bar\psi_A{\cal{S}}^{-1}\not\nabla^\prime({\cal{S}}U_A\psi_A)
\ee

Then the interaction terms are invariant if
\be
\not\nabla^\prime = {\cal{S}}\not\nabla{\cal{S}}^{-1}
\ee
and by (6.8)
\begin{eqnarray}
\not{\cal{W}}^\prime &=& {\cal{S}}\not{\cal{W}}{\cal{S}}^{-1} +
{\cal{S}}\not\partial{\cal{S}}^{-1} \nonumber \\
&=& (Ss)\not{\cal{W}}(s^{-1}S^{-1}) + S\not\partial S^{-1}
\end{eqnarray}
since $s$ is global.  
The field strengths, given by (6.9), transform as
follows:
\begin{eqnarray}
{\cal{W}}^\prime_{\mu\lambda} &=& (\nabla^\prime_\mu,\nabla^\prime_\lambda) \\
&=& {\cal{S}}(\nabla_\mu,\nabla_\lambda){\cal{S}}^{-1} \\
&=& {\cal{S}} {\cal{W}}_{\mu\lambda}{\cal{S}}^{-1}
\end{eqnarray}
and the (standard non-Abelian) field invariant will transform as
\be
\Tr~({\cal{W}}_{\mu\lambda}{\cal{W}}^{\mu\lambda})^\prime = \Tr {\cal{S}}
{\cal{W}}_{\mu\lambda}{\cal{W}}^{\mu\lambda}{\cal{S}}^{-1}
\ee
where the trace is on the $t$ matrices.  Then
\begin{eqnarray}
\Tr ~{\cal{S}}({\cal{W}}_{\mu\lambda}{\cal{W}}^{\mu\lambda}){\cal{S}}^{-1} &=&
\Tr~sS({\cal{W}}_{\mu\lambda}{\cal{W}}^{\mu\lambda}) S^{-1}s^{-1} \\
&=& s\cdot\Tr~S({\cal{W}}_{\mu\lambda}{\cal{W}}^{\mu\lambda})
S^{-1}\cdot s^{-1} \\
&=& s\cdot\Tr~{\cal{W}}_{\mu\lambda}{\cal{W}}^{\mu\lambda}\cdot
s^{-1}
\end{eqnarray}
since the matrix elements of $t_k$ are numerically valued and 
therefore the matrix elements of $S(\vec t)$ commute with the
matrix elements of ${\cal{W}}^{\mu\lambda}(\vec\tau)$.

To reduce (9.20) let us take the trace of (8.6) as follows:
\be
{\rm Tr}~{\cal{W}}_{\mu\lambda}{\cal{W}}^{\mu\lambda} =
W^m_{\mu\lambda}W^{p\mu\lambda}({\rm Tr}~\tau_m\tau_p)
{\cal{D}}_m{\cal{D}}_p+2\hat W^m_{\mu\lambda}
\hat{\cal{W}}^{p\mu\lambda}{\cal{D}}_m{\cal{D}}_p
\ee

Then
\be
{\cal{S}}({\rm Tr}~{\cal{W}}_{\mu\lambda}{\cal{W}}^{\mu\lambda})
{\cal{S}}^{-1} = W^m_{\mu\lambda}W^{p\mu\lambda}
({\rm Tr}~t_mt_p)s{\cal{D}}_m{\cal{D}}_ps^{-1} +
2\hat W^m_{\mu\lambda}\hat W^{p\mu\lambda}s{\cal{D}}_m{\cal{D}}_p
s^{-1}
\ee

By (7.20)-(7.23) the first term on the right side of (9.21) 
vanishes unless
\[ 
(m,p) = (\pm,\mp) \quad  {\mbox{or}} \quad (m,p) = (3,3)
\]  
Hence the first term vanishes unless
\be
s{\cal{D}}_m{\cal{D}}_ps^{-1}= s{\cal{D}}_\pm{\cal{D}}_\mp
s^{-1} 
\ee
or
\be
s{\cal{D}}_m{\cal{D}}_ps^{-1} = s{\cal{D}}_3{\cal{D}}_3 s^{-1}
\ee
But ${\cal{D}}_3{\cal{D}}_3$ as well as ${\cal{D}}_\pm{\cal{D}}_\mp$
carries zero $Q_a$ and $Q_b$ charge and is therefore invariant 
under $s$ transformations.  The second term is invariant for the
same reason since $\hat W_{\mu\lambda}^{~~m}\hat W^{\mu\lambda p}$ 
vanishes by (8.5) unless
$(m,p) = (3,3)$.  Therefore
\be
{\cal{S}}({\rm Tr}~{\cal{W}}_{\mu\lambda}{\cal{W}}^{\mu\lambda})
{\cal{S}}^{-1} = {\rm Tr}~{\cal{W}}_{\mu\lambda}
{\cal{W}}^{\mu\lambda}
\ee
i.e. the standard field invariant remains invariant when
constructed out of the modified vector potential and transformed
according to the new gauge group.

\vskip.5cm

\section{The Higgs Sector.}
\no (a) The Vector Masses.

The neutral couplings in the knot model are by (6.6)
\be
ig\not W_3\tau_3 + ig_oW_o\tau_o
\ee
Introducing the physical fields ($A$ and $Z$) in the standard way
we have
\begin{eqnarray}
W_o &=& A\cos\theta-Z\sin\theta \nonumber \\
W_3 &=& A\sin\theta + Z\cos\theta
\end{eqnarray}
Then (10.1) becomes
\be
A{\cal{A}} + Z{\cal{Z}}
\ee
where
\begin{eqnarray}
{\cal{A}} &=& i(g\tau_3\sin\theta + g_o\tau_o\cos\theta) \\
{\cal{Z}} &=& i(g\tau_3\cos\theta - g_o\tau_o\sin\theta)
\end{eqnarray}
Since there is no interaction between photons and neutrinos one
may write, by (10.3) and (6.3)
\be
(\bar D_\nu^{3/2}~ 0){\cal{A}} \left(
\begin{array}{c}
D_\nu^{3/2} \\ 0 
\end{array} \right)
= \bar D_\nu^{3/2} {\cal{A}}_{\frac{1}{2}\frac{1}{2}}D_\nu^{3/2} = 0
\ee
According to (10.4) the preceding equation is satisfied by
\be
g(\tau_3)_{\frac{1}{2}\frac{1}{2}}\sin\theta +
g_o(\tau_o)_{\frac{1}{2}\frac{1}{2}}\cos\theta = 0
\ee
and by (10.5)
\be
{\cal{Z}} = ig \frac{1}{\cos\theta} (\tau_3)
\ee
Then the covariant derivative of a neutral state is
\be
\nabla_\mu = \partial_\mu + ig
\left[W_+\tau_++W_-\tau_-+\frac{Z\tau_3}{\cos\theta}\right]
\ee
Denote the neutral Higgs scalar by
\be
\phi = \rho D_\nu|0\rangle
\ee
where $D_\nu$ is defined in (5.13) and is the
neutral trefoil, namely (-3,2), carrying the representation
$D^{3/2}_{-\frac{3}{2}\frac{3}{2}}$.

We now replace the kinetic energy term of the neutral Higgs of the
standard model by
\begin{eqnarray}
& &\frac{1}{2}~{\rm Tr}(\overline{\nabla_\mu\varphi}
\nabla^\mu\varphi) \nonumber \\
& &=\frac{1}{2}~{\rm Tr}\langle 0|\bar D_\nu \nonumber \\
& &\mbox{}~~\times\left[\partial_\mu\rho
\partial^\mu\rho+g^2\rho^2[W_+^\mu W_{+\mu}\bar\tau_+\tau_+ 
+ W_-^\mu W_{-\mu}\bar\tau_-\tau_-+\frac{Z^\mu Z_\mu}{\cos^2\theta}
\bar\tau_3\tau_3]\right]D_\nu|0\rangle \\
& &=I~\partial_\mu\rho\partial^\mu\rho+g^2\rho^2
\left[I_{++}W_+^\mu W_{+\mu}+I_{--}W_-^\mu W_{-\mu} +
\frac{I_{33}}{\cos^2\theta}~Z^\mu Z_\mu\right]
\end{eqnarray}
where
\begin{eqnarray}
I~~ &=& \frac{1}{2}~{\rm Tr}\langle 0|\bar D_\nu D_\nu|0\rangle 
\nonumber \\
I_{++} &=& \frac{1}{2}~{\rm Tr}\langle 0|\bar D_\nu\bar\tau_+
\tau_+D_\nu|0\rangle \\
I_{--} &=& \frac{1}{2}~{\rm Tr}\langle 0|\bar D_\nu\bar\tau_-
\tau_-D_\nu|0\rangle \nonumber \\
I_{33} &=& \frac{1}{2}~{\rm Tr}\langle 0|\bar D_\nu\bar\tau_3
\tau_3D_\nu|0\rangle \nonumber
\end{eqnarray}
To agree with the masses predicted by the standard theory (10.12)
must be reduced to the following
\be
\partial_\mu\bar \rho\partial^\mu\bar\rho+g^2\bar\rho^2
\left[W_+^\mu W_{+\mu}+W_-^\mu W_{-\mu} + \frac{1}{\cos^2\theta}
Z^\mu Z_\mu\right]
\ee
where
\be
\bar\rho = I^{1/2} \rho
\ee
To achieve this reduction we impose the following relations:
\be
\frac{I_{kk}}{I} = 1 \qquad k = (+,-,3)
\ee
or
\be
\frac{{\rm Tr}\langle 0|\bar D_\nu(\bar\tau_k\tau_k)
D_\nu|0\rangle}
{{\rm Tr}\langle 0|\bar D_\nu D_\nu|0\rangle} = 1 \qquad
k = (+,-,3)
\ee
By (6.7) and (10.17) we have
\be
|c_k|^{-2} = \frac{\langle 0|\bar D_\nu(\bar{\cal{D}}_k
{\cal{D}}_k)D_\nu|0\rangle}
{\langle 0|\bar D_\nu D_\nu|0\rangle} \qquad k = (+,-,3)
\ee
In (6.7) the coefficients $(c_\pm,c_3)$ were introduced as arbitrary
functions.  They are now fixed by (10.18) as definite functions of
$q$ and $\beta$.  Here the ${\cal{D}}_k$ are given by (5.14).

One finds by (5.13) and (5.14)
\begin{eqnarray}
|c_-|^{-2} &=& \frac{\langle 0|b^3(\bar b^3\bar a^3)(a^3b^3)
\bar b^3|0\rangle}
{\langle 0|b^3\bar b^3|0\rangle} \\
&=& |\beta|^6\langle 0|\bar a^3a^3|0\rangle \\
&=& |\beta|^6 \prod^3_{t=1} (1-q_1^{2t}|\beta|^2) \\
|c_+|^{-2} &=& |\beta|^6\langle 0|a^3\bar a^3|0\rangle \\
&=& |\beta|^6\prod^2_{t=0}(1-q^{2t}|\beta|^2) \\
|c_3|^{-2} &=& \frac{\langle 0|b^3\bar{\cal{D}}_3{\cal{D}}_3
\bar b^3|0\rangle}{\langle 0|b^3\bar b^3|0\rangle} \\
&=& \langle 0|\bar{\cal{D}}_3{\cal{D}}_3|0\rangle \\
&=& \langle 0|[f(\bar bb)]^2|0\rangle \nonumber \\
&=& [f(|\beta|^2)]^2 
\end{eqnarray}
where
\be
f(\bar bb) = D^3_{oo}
\ee
Here $D^3_{oo}$ is the polynomial computed from (3.1).

$c_o$ is not determined by (10.18) but by (10.7).

One finds by (10.7)
\be
\frac{c_o}{c_3} = - \frac{\langle 0|D^3_{oo}|0\rangle}
{\langle 0|D^3_{-11}|0\rangle}
\ee
where the Weinberg relation
\be
\tan\theta = \frac{g_o}{g}
\ee
has been assumed.

\vskip.3cm

\no (b) The Fermion Masses.

In the mass term of (2.1) one has
\be
\bar L\varphi R + \bar R\bar\varphi L
\ee

In the present model this term would be invariant if $L$ is chosen 
to be an external left doublet $\otimes$ an internal $SU_q(2)$
trefoil, and $R$ is chosen to be an external right singlet $\otimes$
internal $SU_q(2)$ singlet, while the Higgs $\varphi$ is chosen
to be an external doublet $\otimes$ internal trefoil adjoint to the
left trefoil.  One would then find for the mass of the n$^{\rm th}$
fermion as an excited state of the $(w,r)$ soliton the following:
\be
m_n(w,r) = \rho(w,r) \langle n|\bar D^{3/2}_{\frac{w}{2}\frac{r+1}
{2}}D^{3/2}_{\frac{w}{2}\frac{r+1}{2}}|n\rangle
\ee
Here $\bar D$ represents the internal state of the soliton and
$D$ the internal state of a Higgs particle.  This speculative
expression is discussed in the first two references.

\vskip.5cm

\section{Remarks.}

We have not studied the perturbative formulation of the theory in higher
orders and therefore do not know whether standard renormalization
procedures can be adapted to the different structure 
coefficients of this formalism,
or whether different procedures for dealing with the higher order
corrections are required.  The theory is gauge invariant.

The purpose of the present work is to provide a field theoretic basis
for our earlier phenomenology.  The main motivation remains the
possibility of providing a substructure for standard theory by 
replacing point particles by solitons.

\ve

The possiblity that the elementary particles are knots has been raised
by many authors.  A particular model related to the Skyrme soliton has
been described by Fadeev and Niemi.$^4$  The present work should be
regarded as a model independent description of the elementary
particles as knots in the context of electroweak theory.

\vskip.5cm

\no {\bf References.}

\vskip.3cm

\begin{enumerate}
\item R. J. Finkelstein, Int. J. Mod. Phys. A{\bf 20}, 487
(2005).
\item R. J. Finkelstein and A. C. Cadavid, Int. J. Mod.
Phys. A{\bf 21}, 4269 (2006).
\item R. J. Finkelstein, ``Trefoil Solitons, Elementary 
Fermions, and $SU_q(2)$", hep-th/0602098.
\item L. Fadeev and Antti J. Niemi, ``Knots and Particles",
hep-th/9610193.
\end{enumerate}

\end{document}